\documentclass[11pt]{article}
\usepackage{amsfonts,amsmath,latexsym,color,epsfig}

\def\S{\mathcal{S}}
\def\R{\mathcal{R}}

\def\eps{\varepsilon}
\def\qed{\ifvmode\mbox{ }\else\unskip\fi\hskip 1em plus 10fill$\Box$}

\long\def\ignore#1{}

\def\rc{\advance\leftskip by 0pt plus 40em\rightskip=\leftskip
\parfillskip=0pt
\spaceskip=.3333em \xspaceskip=.5em \pretolerance=9999
\tolerance=9999 \hyphenpenalty=9999 \exhyphenpenalty=9999}

\title{Tight lower bounds for the size of epsilon-nets}

\author{J\'anos Pach\thanks{EPFL, Lausanne and R\'enyi Institute, Budapest. Supported by NSF Grant CCF-05-14079, by OTKA, and by Swiss National Science Foundation Grant 200021-125287/1.
Email: {\tt pach@cims.nyu.edu}} \and G\'abor
Tardos\thanks{Department of Computer Science, Simon Fraser
University, Burnaby and R\'enyi Institute, Budapest. Supported by NSERC grant 329527, OTKA grants T-046234, AT-048826, and NK-62321, and by the Bernoulli Center at EPFL. Email: {\tt tardos@cs.sfu.edu}}}

\date{}

\begin{document}

\maketitle

\begin{abstract}
According to a well known theorem of Haussler and Welzl (1987), any range space of bounded VC-dimension admits an $\eps$-net of size $O\left(\frac{1}{\eps}\log\frac1{\eps}\right)$. Using probabilistic techniques, Pach and Woeginger (1990) showed that there exist range spaces of VC-dimension 2, for which the above bound can be attained. The only known range spaces of small VC-dimension, in which the ranges are geometric objects in some Euclidean space and the size of the smallest $\eps$-nets is superlinear in $\frac1{\eps}$, were found by Alon (2010). In his examples, the size of the smallest $\eps$-nets is $\Omega\left(\frac{1}{\eps}g(\frac{1}{\eps})\right)$, where $g$ is an extremely slowly growing function, closely related to the inverse Ackermann function.

\smallskip

We show that there exist geometrically defined range spaces, already of VC-dimension $2$, in which the size of the smallest $\eps$-nets is $\Omega\left(\frac{1}{\eps}\log\frac{1}{\eps}\right)$.
We also construct range spaces induced by axis-parallel rectangles in the plane, in which the size of the smallest $\eps$-nets is $\Omega\left(\frac{1}{\eps}\log\log\frac{1}{\eps}\right)$. By a theorem of Aronov, Ezra, and Sharir (2010), this bound is tight.

\ignore{

Let $F(\eps)$ denote the smallest number of points that can be selected from the unit square so that every quasi-rectangle (see
section~\ref{quasisubsection} for the definition) of area $\eps$ contains at least one of them. We prove $F(\eps)=\Theta\left(\frac{1}{\eps}\log\frac{1}{\eps}\right)$,
which settles a weaker version of a problem of Danzer.}
\end{abstract}

\section{Introduction}

Let $X$ be a {\em finite} set and let $\mathcal R$ be a system of subsets of an underlying set which contains $X$. In computational geometry, the pair $(X,\mathcal R)$ is usually called a {\em range space}. The elements of $X$ and $\mathcal R$ are said to be the {\em points} and the {\em ranges} of the range space, respectively. Consider a subset $A\subseteq X$. It is called {\em shattered} if for every subset $B\subseteq A$, one can find a range $R_B\in\mathcal R$ with $R_B\cap A=B$. The size of the largest shattered subset of points, $A\subseteq X$, is said to be the {\em Vapnik-Chervonenkis dimension} (or {\em VC-dimension}) of the range
space $(X,\mathcal R)$.
\medskip

In their seminal paper \cite{VaC71}, Vapnik and Chervonenkis proved that, from the point of view of random sampling, all range spaces whose VC-dimensions are bounded by a constant behave very nicely. In particular, for any $\eps>0$, a randomly selected ``small" subset of $X$, whose number of elements depends only on the VC-dimension $d$ and $\eps$, will ``hit" every range containing at least $\eps|X|$ points of $X$, with large probability. A set of points in $X$ with
the property that every range $R\in\mathcal R$ with $|R\cap X|\ge\eps|X|$
contains at least one of its elements is called an {\em $\eps$-net} for the
range space $(X, {\mathcal R})$. Note that these sets are often called {\em strong} $\eps$-nets in the literature, to distinguish them from the so-called {\em weak $\eps$-nets}, which may also contain points from $\cup{\mathcal R}\setminus X$, but must still hit all ranges that contain at least $\eps|X|$ elements of $X$. In this paper, we will consider only strong $\eps$-nets, apart from some remarks in the last section.

\medskip

The ideas of Vapnik and Chervonenkis have been adapted by Haussler and Welzl \cite{HaW87} to show that the minimum number $f=f_d(\eps)$ such that every range space of VC-dimension $d$ admits an $\eps$-net of size at most $f$ satisfies $f_d(\eps)=O\left(\frac{d}{\eps}\log\frac{d}{\eps}\right)$. They
asked whether the logarithmic factor can be removed in this formula. Pach and Woeginger ~\cite{PaW90} proved that while
$f_1(\eps)=\max(2,\lceil\frac1\eps\rceil-1)$, the logarithmic factor is needed for every $d\ge 2$. Moreover, it was shown by Koml\'os et al.~\cite{KoPW92, PaA95}) that for any $d\ge 2$,
$$(d-2+\frac{1}{d+2}+o(1))\frac{1}{\eps}\log\frac{1}{\eps} \leq f_d(\eps)\leq (d+o(1))\frac{1}{\eps}\log\frac{1}{\eps},$$
as $\eps$ tends to 0. (Here $\log$ denotes the natural logarithm.)

\medskip

Haussler and Welzl discovered that the above results apply to many
geometrically defined range spaces. Roughly speaking, the VC-dimension is bounded by a constant for any set of ranges with bounded {\em description complexity}, that is if the ranges can be described in terms of a bounded number of parameters. This observation has far reaching consequences. The construction
of small epsilon-nets has become one of the most powerful general techniques in computational geometry (see \cite{Ch00, EvRS05}).

\medskip

In a number of basic geometric scenarios it was possible to improve on the above bounds. For instance, for any finite set of points in the plane, one can find an $\epsilon$-net of size linear in $1/\eps$, where the ranges are half-planes, translates of a convex polygon, disks or certain kind of pseudo-disks. Similar results hold in three-dimensional space for half-space ranges \cite{PaW90, MaSW90, Ma92, PyR08}. We state two results here.

\medskip
\noindent{\bf Theorem A.} (Matou\v sek, Seidel, Welzl \cite{MaSW90, Ma92}) {\em All range spaces $(X,\mathcal R)$, where $X$ is a finite set of points in ${\mathbb{R}}^3$ and $\mathcal R$ consists of half-spaces, admit $\eps$-nets of size $O(1/\eps)$.}
\medskip

\noindent{\bf Theorem B.} (Aronov, Ezra, Sharir \cite{ArES10})
{\em All range spaces $(X,\mathcal R)$, where $X$ is a finite set of points in ${\mathbb{R}}^2$ (or ${\mathbb{R}}^3$) and $\mathcal R$ consists of axis-parallel rectangles (boxes), admit $\eps$-nets of size $O\left(\frac{1}{\eps}\log\log\frac1{\eps}\right)$.}

\medskip

\noindent Aronov et al. have also established a similar result for ``fat" triangular ranges in the place of axis-parallel rectangles. For weak $\eps$-nets, Ezra \cite{Ez10} extended Theorem~B to higher dimensions.

\smallskip

In algorithmic applications, it is often natural to consider the dual range space, in which the roles of points and ranges are
swapped \cite{BrG95, PaA95}. Given a finite family $\mathcal R$ of ranges in ${\mathbb{R}}^m$, the {\em dual range space induced} by them is defined as a set system (hypergraph) on the underlying set $\mathcal R$, consisting of the
sets ${\mathcal R}_x:=\{R \mid x\in R\in{\mathcal R}\}$, for all $x\in {\mathbb{R}}^m$. (Note that ${\mathcal R}_x$ and${\mathcal R}_y$ may coincide for $x\neq y$.) It is easy to see that if the VC-dimension of the range space $(X, {\mathcal R})$ is less than $d$ for every $X\subset \mathbb{R}^m$, then the VC-dimension of the dual range space induced by any subset of $\mathcal R$ is less than $2^d$.

Clarkson and Varadarajan~\cite{ClV07} found a simple and beautiful connection between the complexity of the boundary of the union of $n$ members of $\mathcal R$ and the size of the smallest epsilon-net in the dual range space. If the complexity of the boundary is $o(n\log n)$, then the dual range space admits $\eps$-nets of size
$o\left(\frac{1}{\eps}\log\frac1{\eps}\right)$. This connection has been further explored and improved in \cite{Va09, ArES10}. In particular, it was shown that dual range spaces of ``fat" triangles in the plane admit $\eps$-nets of size
$O\left(\frac{1}{\eps}\log\log\log\frac1{\eps}\right)$.

\medskip

In most range spaces $(X,\mathcal R)$, one can find roughly $1/\eps$ pairwise disjoint ranges $R\in{\mathcal R}$ such that the sets $R\cap X$ are of size at least $\eps |X|$. In these cases, the size of any
$\eps$-net is $\Omega(1/\eps)$. For the last two decades, ``the
prevailing conjecture" was that in ``geometric scenarios," this bound is essentially tight: there always exists an $\eps$-net of size $O(1/\eps)$ (see, e.g., \cite{MaSW90, ArES10}. This conjecture had to be revised after Alon~\cite{Al10} discovered some geometric range spaces of small VC-dimension, in which the ranges are straight lines, rectangles or infinite strips in the plane, and which do not admit $\eps$-nets of size $O(1/\eps)$. Alon's construction is based on the density version of the Hales-Jewett theorem~\cite{HaJ63}, due to Furstenberg and Katznelson~\cite{FuK89, FuK91}, and recently improved in \cite{Po09}. However, his lower bound is only barely
superlinear: $\Omega\left(\frac{1}{\eps}g(\frac{1}{\eps})\right)$, where $g$ is an extremely slowly growing function, closely related to the inverse Ackermann function.

\subsection{New lower bounds}

The main aim of this note is to prove that the $O\left(\frac{1}{\eps}\log\frac{1}{\eps}\right)$
general upper bound for the size of the smallest $\eps$-nets in range spaces of bounded dimension is tight even in simple geometric scenarios.

\smallskip


Our first theorem claims that there exist dual range spaces induced by finite families of axis-parallel rectangles in which the size of the smallest $\eps$-nets is $\Omega\left(\frac{1}{\eps}\log\frac{1}{\eps}\right)$.
More precisely, we have the following.

\medskip
\noindent {\bf Theorem 1.} {\em For any $\eps>0$ and for any sufficiently large integer $n>n_0(\eps)$, there exists a dual range space $\Sigma^*$ of VC-dimension 2, induced by $n$ axis-parallel rectangles in ${\mathbb{R}}^2$, in which the minimum size of an $\eps$-net is at least
$C\frac{1}{\eps}\log\frac{1}{\eps}$. Here $C>0$ is an absolute constant.}

\medskip

From Theorem 1 it is not hard to deduce the following results for primal range spaces.

\medskip
\noindent{\bf Theorem 2.} {\em For any $\eps>0$ and for any sufficiently large integer $n>n_0(\eps)$, there exists a (primal) range space $\Sigma=(X,\mathcal R)$ of VC-dimension 2, where $X$ is a set of $n$ points in ${\mathbb{R}}^4$, $\mathcal R$ consists of axis-parallel boxes with one of their vertices at the origin, and in which the size of the smallest $\eps$-net is at least $C\frac{1}{\eps}\log\frac{1}{\eps}$. Here $C>0$ is an absolute constant. }

\medskip
\noindent{\bf Theorem 3.} {\em For any $\eps>0$ and for any sufficiently large integer $n>n_0(\eps)$, there exists a (primal) range space $\Sigma=(X,\mathcal R)$ of VC-dimension 2, where $X$ is a set of $n$ points in ${\mathbb{R}}^4$, $\mathcal R$ consists of half-spaces, and in which the size of the smallest $\eps$-net is at least $C\frac{1}{\eps}\log\frac{1}{\eps}$. Here $C>0$ is an absolute constant.}
\medskip

Theorems 2 and 3 show that Theorems B and A cannot be generalized to 4-dimensional space. It also follows, by a standard duality argument, that there exist {\em dual} range spaces induced by half-spaces in ${\mathbb{R}}^4$, for which the size of the smallest $\eps$-net is $\Omega\left(\frac{1}{\eps}\log\frac{1}{\eps}\right)$.

Our next result shows that Theorem B of Aronov, Ezra, and Sharir is tight.

\medskip

\noindent {\bf Theorem 4.} {\em For any $\eps>0$ and for any sufficiently large integer $n>n_0(\eps)$, there exists a (primal) range space $\Sigma=(X,\mathcal R)$, where $X$ is a set of $n$ points in the plane, $\mathcal R$ consists of axis-parallel rectangles, and
in which the size of the smallest $\eps$-net is at least
$C\frac{1}{\eps}\log\log\frac{1}{\eps}$. Here $C>0$ is an absolute constant.}
\medskip

The VC-dimension of the family of {\em all} axis-parallel
rectangles in the plane is 4. However, it is easy to verify that the VC-dimension of the range spaces used for the proof of Theorem~4 is only at most 3. In the full version of this paper, we also outline a somewhat different approach to prove the existence of range spaces of VC-dimension 2 that satisfy the conditions in Theorem~4.

The proofs of Theorems 1 and 4 are based on two constructions from
\cite{PaT10} and \cite{ChPS09}, related to hypergraph coloring problems.

\ignore{
\subsection{Quasi-rectangles in the unit square}\label{quasisubsection}

An old problem of Danzer and Rogers (often attributed to Danzer; see, e.g., \cite{BeC87}, p. 285) is the following. Given $\eps>0$, what is the size of the smallest set of points with the property that every compact convex subset of the unit square of area $\eps$ contains at least one of them. Denoting this minimum by
$f(\eps)$, we clearly have $f(\eps)=\Omega(1/\eps)$. The question is whether $f(\eps)=O(1/\eps)$ holds.

This problem can be regarded as a continuous version of the $\eps$-net problem in an infinite range space, where the ground set $X$ is the unit square, the ranges are compact convex subsets of $X$, and we want to hit every range $R$ with $|R\cap X|=|R|\geq \eps|X|$, where $|.|$ stands for the Lebesgue measure (area).

The area of the largest rectangle contained in a plane convex set $R$ is at least half of the area of $R$ \cite{Ra57}. Thus, in order to ``stab" all plane convex sets of area $\eps$, it is sufficient to find an $\eps/2$-net for rectangles. The family of rectangles has bounded VC-dimension. Therefore, the theorem of Haussler and Welzl implies that $f(\eps)=O\left(\frac{1}{\eps}\log\frac{1}{\eps}\right)$.

We cannot decide whether this bound is tight. However, if we slightly enlarge the family of rectangles, by also including ``quasi-rectangles," then we can prove that the logarithmic overhead in the solution of the corresponding problem is really necessary.

\medskip

A rectangle is a region swept out by a line segment $s$ moving orthogonally to itself. If we continuously translate $s$ almost orthogonally to itself, without rotating it, so that the angle between $s$ and the trajectory of its center always remains between $90-\delta$ and $90+\delta$ degrees for a fixed small $\delta>0$, then we call the resulting region a {\em quasi-rectangle}. To be concrete, set $\delta=1^{\circ}$. Note that the motion of the segment $s$ is supposed to be monotone in the direction orthogonal to it, so that the segment is not allowed to turn back. Therefore, the area of a quasi-rectangle is equal to the length of $s$ multiplied by the distance it traveled in the direction orthogonal to $s$.

A quasi-rectangle is not necessarily convex, but it is ``almost"
convex. Although the VC-dimension of the family of quasi-rectangles is unbounded, it is not hard to see that all quasi-rectangles of area $\eps$ inside the unit square can be stabbed by
$O\left(\frac{1}{\eps}\log\frac{1}{\eps}\right)$ points.
Our next theorem shows that this bound is tight up to a constant factor.

\medskip
\noindent{\bf Theorem 5.}
{\em For any $\eps>0$, let $F(\eps)$ denote the smallest number of points that
that every quasi-rectangle inside the unit square of area $\eps$ contains at
least one of them. We have
$F(\eps)=\Theta\left(\frac{1}{\eps}\log\frac{1}{\eps}\right)$.}
}

\subsection{Organization}
In Section 2, we present the proofs of Theorems 1, 2, and 3, based on an explicit construction of systems of axis-parallel rectangles, described in \cite{PaT10}. Section 3 contains a similar proof of Theorem 4, based on randomized construction from Chen et al.~\cite{ChPS09}.
In the final section, we make some concluding remarks and mention some open problems.

\section{Boxes and half-spaces---Proofs of Theorems 1-3}

Theorems 2 and 3 are corollaries of Theorem 1, so we start with the proof of Theorem 1. The proof is based on an explicit construction of systems of rectangles, presented in \cite{PaT10}. In order to describe this construction, we have to introduce some notations.

\smallskip

For any two integers $c\ge2$ and $k\ge0$, let
$[c]:=\{0,1,\ldots,c-1\}$ and let $[c]^k$ stand for the set of
strings of length $k$ over the alphabet $[c]$. For $x\in[c]^k$,
let $x_j$ denote the $j$th digit of $x$ ($1\le j\le k$), so that
we have $x=x_1\ldots x_k$. For notational convenience we write
$x_0=0$. Expanding $x$ as a $c$-ary fraction, we obtain a number
$\overline{x}:=\sum_{j=1}^kx_j/c^j$. Let $\theta$ denote the empty
string so that $\overline\theta=0$.

Let $c\ge2$ and $d\ge1$ be integers. For any $0\le k\le d$,
$u\in[c]^k$, and $v\in[c]^{d-k}$, define an open axis-parallel
rectangle $R^k_{u,v}$ in the unit square, as follows:
$$R^k_{u,v}:=(\overline u,\overline u+c^{-k})\times(\overline
v,\overline v +c^{k-d})$$
and consider the family
$$\R=\R(c,d)=\left\{R^k_{u,v} \ \mid \ 0\le k\le d,\; u\in[c]^k,\; v\in[c]^{d-k},\; u_k=v_{d-k}\right\}.$$
Clearly, we have $|\R|=(d+1)c^{d-1}$. Finally, let
$\Sigma=\Sigma(c,d)$ be the (infinite) range space $({\mathbb R}^2,\R)$ and let $\Sigma^*=\Sigma^*(c,d)$ denote its dual. That is, the underlying set of $\Sigma^*$ is $\R=\R(c,d)$, and its
ranges (hyperedges) are all sets of the form $\{R\in\R \mid x\in R\}$ for some $x\in{\mathbb{R}}^2$.


\medskip

The most important property of our construction is the following.

\medskip
\noindent{\bf Lemma 2.1.} {\em Let $d\ge1$, $r\ge2$, $c\ge3$ and let $\Sigma^*=\Sigma^*(c,d)$ denote the dual range space defined above. If a subset $I\subseteq\R(c,d)$ contains no $r$-element range (hyperedge) of $\Sigma^*$, then we have
$$|I|\le(r-1)\frac{c-1}{c-2}c^{d-1}.$$}

\medskip

In \cite{PaT10}, we established the slightly weaker bound
$|I|\le\frac{c^{d-1}}{\frac1{r-1}-\frac1{c-1}}$, for any $c>r$. The main focus of that paper was the case $r=2$, in which the two bounds
coincide.

\medskip

\noindent{\bf Proof of Lemma 2.1.} To explain the proof, first we have to sketch the original argument from \cite{PaT10}. Two distinct rectangles $R,R'\in\R$ are called {\em siblings} if
$R=R^k_{u,v}$, $R'=R^k_{u',v'}$, where $u$ and $u'$ differ only in their last digit, and the same is true for $v$ and $v'$. Clearly, for $1\le k<d$, the rectangles of the form $R^k_{u,v}\in\R$ fall into groups, each consisting of $c$ siblings. For $k=0$ and $k=d$, $R^k_{u,v}$ has no sibling. A rectangle $R\in\R$ is called {\em bad} if $R\notin I$, but for each of its $c-1$ siblings we have $R'\in I$. Let $B$ denote the set of bad rectangles.

Using the assumption that $I\subseteq\R$ contains no $r$-element range of $\Sigma^*$, we proved in \cite{PaT10} that
\begin{equation}\label{x}
|I|\le(r-1)|B|+(r-1)c^{d-1}.
\end{equation}
Comparing (\ref{x}) to the trivial inequality $|B|\le|I|/(c-1)$, we obtain the weaker bound
$|I|\le\frac{c^{d-1}}{\frac1{r-1}-\frac1{c-1}}$.

Now we choose a different strategy to deal with bad rectangles. For every $R\in B$, we pick one of the $c-1$ siblings of $R$ and remove it from the set $I$. Since the resulting set $I'\subseteq I$ contains no $r$-element range in $\Sigma^*$, we can apply inequality (\ref{x}) to $I'$. By the construction of $I'$, the corresponding set $B'$ of bad rectangles in $\R$ is empty, so that we obtain
$$|I'|\le(r-1)c^{d-1}.$$
Comparing this inequality to $|I'|=|I|-|B|\ge\frac{c-2}{c-1}|I|$, the lemma follows. $\Box$

\medskip

\noindent{\bf Lemma 2.2.} {\em Both $\Sigma$ and $\Sigma^*$ have VC-dimension
2.}

\medskip
Before turning to the proof, we have to introduce a partial order on the family of axis-parallel rectangles in the plane. For any two axis-parallel rectangles $R$ and $R'$, we write $R\prec R'$ if the orthogonal projection of $R$ on the $x$-axis is contained in the orthogonal projection of $R'$ on the $x$-axis, and the
orthogonal projection of $R$ on the $y$-axis contains the orthogonal projection of $R'$ on the $y$-axis. Obviously, this is a partial order.
\medskip

\noindent{\bf Proof of Lemma 2.2.} Clearly, we have VC-dim$(\Sigma^*)$, VC-dim$(\Sigma)\geq 2$.

Observe first that no rectangle in $\R$ contains a vertex of any other rectangle in its interior. This implies that any two intersecting rectangles in $\R$ are comparable by $\prec$.

Assume for contradiction that $\Sigma$ or $\Sigma^*$ has VC-dimension 3 or more. In either case, the existence of a shattered 3-element set would mean that there are three distinct points $p_1$, $p_2$, and $p_3$ in the plane and three rectangles $R_1,R_2,R_3\in\cal R$ with
$\{p_1,p_2,p_3\}\setminus R_i=\{p_i\}$ for $i=1,2,3$. The rectangles $R_i$ pairwise intersect, and hence must be linearly ordered by $\prec$. Suppose without loss of generality $R_1\prec R_2\prec R_3$. Then $R_1\cap R_3\subseteq R_2$, contradicting our assumption that $p_2$ is contained in the left-hand side but not in the right. $\Box$

\medskip

\noindent{\bf Proof of Theorem 1.} Let $0<\eps<2^{-6}$, and set
$r=\lceil\log\frac1\eps/6\rceil\ge 2$, $c=4$, and $d=3r-4$, where $\log$ denotes the binary logarithm. Consider the dual
range space $\Sigma^*=\Sigma^*(c,d)$. The number of rectangles in this range space is $|\R|=|\R(c,d)|=(d+1)c^{d-1}$, and by Lemma~2.2, the VC-dimension of $\Sigma^*$ is 2.

Let $\S\subseteq\R$ be an $\eps$-net in $\Sigma^*$, that is, a set of rectangles with the property that any point of the plane which is covered by at least $\eps |\R|$ members of $\R$ is contained in an element of $\S$. Notice that with our choice of parameters we have $\eps|\R|<r$, hence the rectangles in $\R\setminus \S$ cannot induce any $r$-element range (hyperedge) in $\Sigma^*$. Applying Lemma~2.1 with $I=\R\setminus\S$, we obtain that

$$|\R\setminus\S|\le(r-1)\frac{c-1}{c-2}c^{d-1}={(d+1)c^{d-1}\over 2}={|\R|\over 2}.$$

This yields that

$$|\S|\ge{|\R|\over 2}=\frac1\eps\cdot\frac r2\ge\frac{\frac1\eps\log\frac1\eps}{12}.$$

So far our examples may appear quite special, because for every $\eps$, we have defined only one particular space $\Sigma^*$, consisting of
$O\left(\frac{1}{\eps}\log\frac{1}{\eps}\right)$
rectangles. However, from this small example we can easily construct
arbitrarily large ones, as follows. Keep $r$ and the corresponding $\eps$ fixed, and choose a large integer $t$. Replace each rectangle $R\in\R$ by a chain of rectangles $R_1\prec R_2\prec\cdots\prec R_t$, where $\prec$ denotes the ordering relation defined after Lemma~2.2, and each $R_i$ differs only very little from $R$. Note that the dual range space $\Sigma^*$, as well as the corresponding primal space have VC-dimension 2 by Lemma~2.2. It is not difficult to see that if the difference between (the coordinates of) the new rectangles $R_i$ and the original rectangle $R\in\R$ is small enough, then the VC-dimension of the dual range space $\Sigma^*_t$ induced by the resulting family of rectangles ${\R}_t$, as well as the VC-dimension of the ``primal" space $\Sigma_t=({\mathbb R}^2,\R_t)$, remains 2.

We have $|{\R}_t|=t|\R|$, and the size of the smallest $\eps$-net for $\Sigma^*_t$ is at least as large as it was in $\Sigma^*$. Suppose to the contrary that there is a smaller set $\S'$ of rectangles in ${\R}_t$ that form an $\eps$-net in $\Sigma^*_t$. Let ${\S}''$ be the set of rectangles in $\R$ that were replaced by the elements of $\S'$. Since $|{\S}''|\leq |\S'|$, the rectangles in ${\S}''$ do not form an $\eps$-net in $\Sigma^*$. Thus, there is a point in the plane contained in at least $\eps|\R|$ elements of $\R$, which is not covered by any element of ${\S}''$. We can choose such a point lying
not too close to the boundaries of the rectangles in $\R$, and then it is contained in at least $t\eps|\R|=\eps|{\R}_t|$ elements
of ${\R}_t$, none of which belongs to $\S'$, a contradiction. $\Box$

\medskip

\noindent{\bf Proof of Theorem 2.} The statement follows from Theorem 1 by a standard duality argument. We assume without loss of generality the the rectangles are closed and lie in the first quadrant of the plane. We assign to each rectangle $R=[x_1,x_2]\times[y_1,y_2]$ the point $p(R)=(x_1,1/{x_2},y_1,1/{y_2})\in{\mathbb{R}}^4$. Now a point $q=(a,b)$ of the first quadrant lies in $R$ if and only if
$x_1\le a\le x_2$ and $y_1\le b\le y_2$, that is, if and only if the point $p(R)$ is contained in the 4-dimensional box
$$B(q)=[0,a]\times[0,1/a]\times[0,b]\times[0,1/b].\;\;\;\;\;\Box$$
\medskip

Theorem 3 is an immediate corollary of Theorem 2 and the following lemma.

\medskip

\noindent{\bf Lemma 2.3.} {\em Let $P$ be a finite set of points in the positive orthant of ${\mathbb{R}}^d$. To each $p\in P$, we can assign a point $p'$ in the positive orthant of ${\mathbb{R}}^d$ so that the set $P'=\{p' \mid p\in P\}$ satisfies the following condition.

For any axis-parallel box $B\subset{\mathbb{R}}^d$ that contains the origin, there is a half-space $H_B\subset{\mathbb{R}}^d$ which contains the origin and for which
$$\{p' \mid p\in B\cap P\}= P'\cap H(B).$$}

\smallskip

\noindent{\bf Proof.} Let $x_1, x_2,\ldots, x_d$ denote the orthogonal coordinates in ${\mathbb{R}}^d$. Observe that from the point of view of intersections with axis-parallel boxes, the actual values of the coordinates do not matter: we need to know only the order of the $x_i$-coordinates of the points of $P$ for each $i$. For every $i\; (1\le i\le d)$, let $0<\xi_{i,1}<\xi_{i,2}<\xi_{i,3}<\ldots$ denote the sequence of different values of the $x_i$-coordinates of the elements of $P$. Every such sequence is of length at most $|P|$. By rescaling the coordinates if necessary, we can assume that $\xi_{i,j+1}/\xi_{i,j}>d$ holds for every $i$ and $j$.

Consider now an axis-parallel box $B$, which contains the origin and
intersects $P$ in at least one element. We can shrink $B$ if necessary, without changing its intersection with $P$, so that we can suppose without loss of generality that $B$ is of the form
$$B=[0,b_1]\times[0,b_2]\times\ldots\times[0,b_d],$$
where each $b_i$ is equal to $\xi_{ij_i}$ for a suitable $j_i$.

We claim that $B\cap P$ is equal to the intersection of $P$ with the half-space $H(B)$ defined by
$${x_1\over b_1} + {x_2\over b_2} +\ldots+ {x_d \over b_d}\leq d.$$
For every point in $B$, each term of the above sum is at most 1, so that we have $B\subset H(B)$, and hence $B\cap P\subseteq H(B)\cap P$. Suppose now that $p$ is a point of $P$ that does not belong to $B$. Then one of its coordinates, $x_i(p)$, say, is more than $d$ times larger than $b_i$. Therefore, the $i$-th term in the above sum is already larger than $d$, which implies that $p\not\in H(B)$. $\Box$

\section{Proof of Theorem 4}

Theorem~4 is an easy consequence of the following result on a set of  randomly selected points in the unit square. A similar property of random point sets with respect to axis-parallel rectangles was established in Chen et al.~\cite{ChPS09} (see Theorem 9). In their setting, $r$ was a constant, $\eps=r/n$, and it was shown that every $\eps$-net contains all but a very small fraction of point set. Here we allow $r$ to slowly tend to infinity.

\medskip

\noindent{\bf Lemma 3.1.} {\em Let $n>2$, $r=\lceil\log\log n/5\rceil$ be integers, where $\log$ stands for the binary logarithm, and let and $\eps=r/n$. Let $X$ be a set of $n$ randomly and uniformly selected points in the unit square, and let $\R$ denote the family of all axis-parallel rectangles of the form $[j/2^t,(j+1)/2^t)\times[a,b]$, where $j, t$ are nonnegative
integers, and $a<b$ are reals.

Then, with probability tending to 1, the range space $(X,\R)$ does not admit an $\eps$-net of size at most $n/2$.}
\medskip

\noindent{\bf Proof.} We write $[n]$ to denote the index set $\{1,\ldots,n\}$. Let us choose the $y$-coordinates of our random points $p_i$ first, and then enumerate them in the increasing order of their $y$-coordinates. That is, let $p_i=(x_i,y_i)$, where the numbers $y_1<y_2<\cdots<y_n$ are fixed and the $x_i$-s are chosen
uniformly and independently from $[0,1]$. Finally, let $X=\{p_i\mid i\in [n]\}$.

Fix a subset $I\subseteq[n]$ of size at most $n/2$, and let $S_I=\{p_i\mid i\in I\}$. We will prove that the probability that $S_I$ is an $\eps$-net for the range space $(X,\R)$ is very small.

We write each $x_i$ as an infinite binary fraction $0.d_i^{(1)}d_i^{(2)}\ldots$. That is, $x_i=\sum_{i=1}^{\infty}d_i^{(t)}$, where $d_i^{(t)}=0$ or $1$.
The $t$-th {\em truncation} of $x_i$, denoted by $x_i^{(t)}$, is the finite binary fraction $0.d_i^{(1)}d_i^{(2)}\ldots d_i^{(t-1)}$. In particular, we have $x_i^{(1)}=0$.

Choosing $x_i$ uniformly at random can be achieved by selecting all of its binary digits $d_i^{(t)}$ uniformly and independently. This will be done in stages. At stage $t$, we choose $d_i^{(t)}$ for all $i$.

\medskip

Consider now stage $t$ of our selection process for a fixed $t$, $1\le t\le \log(n/r)-1$. Before the selections are made, $x_i^{(t)}$ has been fixed for all $i$. For every $x\in [0,1]$, define
$$H_x=H_x^{(t)}=\{1\le i\le n\mid x_i^{(t)}=x\}.$$
The sets $H_x$ form partition $[n]$ into at most $2^{t-1}$ nonempty parts.

For each $x\in [0,1]$, divide $H_x$ into as many pairwise disjoint {\em intervals} as possible, each containing $r$ elements not in $I$. More precisely, select $\lfloor|H_x\setminus I|/r|\rfloor$ pairwise disjoint sets $H_{x,j}$ of the form $H_{x,j}=\{i\in H_x\mid a_{x,j}\le i\le b_{x,j}\}$ with $|H_{x,j}\setminus I|=r$.

For a given $x$, out of the at least $n/2$ indices in $[n]\setminus I$, there are fewer than $r$ that do not belong to any interval of $H_x$. Using our assumption $t\le\log(n/r)-1$, the total number of indices in $[n]\setminus I$ that belong to some interval $H_{x,j}$ over all $x$ and $j$ is larger than $n-|I|-2^{t-1}r\ge n/4.$ Since each interval contains precisely $r$ such indices, the number of intervals is larger than $n/(4r)$.

We call an interval $H_{x,j}$ {\em bad} if its size is at least $4r$, otherwise is called {\em good}. Any bad interval contains at least $3r$ elements of $I$, so the number of bad intervals is at most $|I|/3r\le n/(6r)$. Consequently, the number of good intervals is at least the total number of intervals minus $n/(6r)$, which is larger than $n/(4r)-n/(6r)=n/(12r)$.

\medskip

Let $G_{x,j}$ be a good interval. With probability $2^{-|G_{x,j}|}>2^{-4r}$ we have $d_i^{(t)}=0$ for
all $i\in G_{x,j}\setminus I$ but $d_i^{(t)}=1$ for all $i\in G_{x,j}\cap I$. If this happens, we say that the interval $G_{x,j}$ {\em fails}. If $G_{x,j}$ fails, then for the rectangle
$R=[x,x+2^{-t})\times[y_{a_{x,j}},y_{b_{x,j}}]$ we have $R\cap X=\{p_i\mid i\in G_{x,j}\setminus I\}$. That is, in this case we have $|R\cap X|=r=\eps n$ and $R\cap S_I=\emptyset$, showing that $S_I$ is not an $\eps$-net for $(X,\R)$.

Notice that at a fixed stage $t\; (1\le t\le \log(n/r)-1)$, all the at least $n/(12r)$ good intervals fail independently, each with probability larger than $2^{-4r}$. We say that $S_I$ {\em survives stage $t$} if none of the intervals fail. We have
$${\mbox{\rm Prob}}[S_I\; {\mbox{\rm survives stage }} t]<
(1-2^{-4r})^{n/(12r)}<2^{-n/(12r2^{4r})}.$$
This inequality holds independently of what happened at the earlier stages, so that
$${\mbox{\rm Prob}}[S_I\; {\mbox{\rm  is an }} \eps{\mbox{\rm -net for }} (X,\R)]<$$
$${\mbox{\rm Prob}}[S_I\; {\mbox{\rm  survives all stages }} t\le \log(n/r)-1)] <2^{-(\log(n/r)-2)n/(12r2^{4r})}.$$

There are fewer than $2^n$ choices for a set $I$ with $|I|\le n/2$. By the union bound, this yields that
$${\mbox{\rm Prob}}[(x,\R)\; {\mbox{\rm  admits an }} \eps{\mbox{\rm -net of size}}\le n/2]<2^{n-(\log(n/r)-2)n/(12r2^{4r})}.$$
The right-hand side of this inequality tends to 0, as $n\rightarrow\infty$. $\Box$

\medskip

\noindent{\bf Proof of Theorem 4.} Consider the random
range space $(X,\R)$ described in Lemma~3.1, where $n$ is so large that the probability that $(X,\R)$ does not admit an $\eps$-net of size at most $n/2$ is positive. Fix an $n$-element set $X$ with this property. Then the minimum size of an $\eps$-net for $(X,\R)$ is larger than
$\frac{n}2=\frac1\eps\cdot\frac r2>{\frac1\eps\log\log\frac1\eps}/10.$

Once we have one example of a range space $\Sigma=(X,\R)$ that admits no small $\eps$-net for a given value of $\eps$, we can create arbitrarily large examples with the same property, by replacing each point $p\in X$ with $t$ new points, very close to $p$. (The same trick was applied in \cite{Al10} and in the proof of Theorem 1.) This completes the proof of Theorem 4.
$\Box$

\medskip

The VC-dimension of the random range space we considered is 3. However, we can also construct a range space of VC-dimension 2, meeting the requirements of Theorem 4.

\section{Concluding remarks}

\noindent{\bf 1.} It was shown in \cite{PaW90} that any range space $(X,\mathcal R)$, where $X$ is a finite point set in the
plane and $\mathcal R$ consists of half-planes, admits $\eps$-nets of size at most $\lceil 2/\eps\rceil -1$, and that this bound is tight up to an additive constant at most 1. The corresponding result on the line is almost trivial. Consequently, Theorem A holds in any dimension $d\leq 3$, and our Theorem 4 shows that it is false for $d>3$.

The epsilon-net problem for half-spaces (containing the origin) is
self-dual. That is, any {\em dual} range space induced by half-spaces in ${\mathbb{R}}^d$ admits an $\eps$-net of size $O(1/\eps)$ if $d\leq 3$, and this statement is false whenever $d>3$.

\medskip

\noindent{\bf 2.} Recall that a {\em weak $\eps$-net} for a range space $(X,\mathcal R)$ is a set of elements of $\cup_{R\in\mathcal R}R$ (not necessarily in $X$) such that every range $R\in \mathcal R$ with $|R\cap X|\ge \eps|X|$ contains at least one of them. In \cite{Ez10}, Ezra proved that if $X$ is any finite set of points in ${\mathbb{R}}^d$ and $\mathcal R$ consists of all axis-parallel boxes, then $(X,\mathcal R)$ admits a weak $\eps$-net of size $O\left(\frac{1}{\eps}\log\log\frac1{\eps}\right)$. This implies that our Theorem~2 cannot be strengthened by requiring that the
constructed range spaces do not admit {\em weak} $\eps$-nets of size smaller than $\frac{1}{\eps}\log\frac{1}{\eps}$, provided that $\eps>0$ is sufficiently small.

It is easy to see that the analogue of Theorem~3 is also false for {\em weak} $\eps$-nets instead of strong ones. Indeed, any finite system of half-spaces in ${\mathbb{R}}^d$ can be hit by $d+1$ points, so that in (primal or dual) half-space range spaces there always exist weak $\eps$-nets of size $O(1)$.

However, we have been unable to decide whether the analogue of  Theorem~4 holds for weak $\eps$-nets in place of strong ones.

\medskip

\ignore{
\noindent{\bf 3.} The definition of quasi-rectangles involves a parameter $\delta$. Recall that a region is called a {\em $\delta$-quasi-rectangle} if it swept out by a segment $s$ translated almost orthogonally to itself with a possibly changing velocity vector that encloses an angle of absolute value at most $\delta$ with the positive normal vector of $s$. As $\delta\rightarrow 0$, a $\delta$-quasi-rectangle resembles more and more a real rectangle.

It is well known that there is a set of $O(1/\eps)$ points in the unit square $N=[0,1]\times[0,1]$ such that every axis-parallel rectangle $R\subset N$ with area at least $\eps>0$ contains at least one of them. It follows from the proof of Theorem 5 that this statement does not remain true for $\delta$-quasi-rectangles, for any fixed $\delta>0$. We have the following result.

\medskip
\noindent{\bf Theorem 5'.} {\em There exists an absolute constant $C>0$ such that for any $\delta, \eps>0$ with $\delta>2\eps$, and for any set of points $S\subset {\mathbb{R}}^2$ with $|S|<C{1\over\eps}\log{\delta\over\eps}$, there is a $\delta$-quasi-rectangle with vertical sides that does not contain any element of $S$. This bound is tight up to the value of the constant $C$.}   \;\;\;\;\;\;\; $\Box$
\medskip
}

\noindent{\bf 3.} Let $X$ be a finite or infinite set and let $\mathcal R$ be a family of ``ranges" of a certain type in $\mathbb{R}^d$  (e.g., lines, balls, half-spaces, axis-parallel boxes). We say that a subfamily $\mathcal S\subset \mathcal R$ forms a {\em $k$-fold covering} of $X$ if every point of $X$ belongs to at least $k$ members of $\mathcal S$. It is an old problem in discrete geometry to decide whether every $k$-fold covering selected from a family $\mathcal R$ can be decomposed into two or more coverings \cite{PaTT09}. For example, it was shown by Gibson and Varadarajan \cite{GiV09} that every $k$-fold covering of the plane with translates of a convex polygon can be decomposed into $\Omega(k)$ coverings.

There is an intimate relationship between epsilon-net problems and problems about decomposition of multiple coverings. If we know that every $k$-fold covering $\mathcal S\subset \mathcal R$ with $|\mathcal S|=n$ splits into at least $ck$ coverings for some absolute constant $c>0$, then one of these coverings contains at most $n/(ck)$ sets. Setting $k=\eps n$, we find a covering consisting of at most $1/(c\eps)$ members of $\mathcal S$. This means that the {\em dual} range space $\Sigma^*$ induced by the members of $\mathcal S$ admits an $\eps$-net of size $O(1/\eps)$. Therefore, if the dual range space does not always admit an $\eps$-net of size $O(1/\eps)$, then it cannot be true that every $k$-fold covering with ranges from $\mathcal R$ splits into $\Omega(k)$ coverings.

In particular, Alon \cite{Al10} proved that there are $n$-element point sets $X\subset {\mathbb{R}}^2$ and straight-line ranges that do not admit $\eps$-nets of size $O(1/\eps)$. The standard duality between points and lines preserves incidences. Switching to the dual, we obtain dual range spaces induced by sets of $n$ lines in the plane that do not admit $\eps$-nets of size $O(1/\eps)$. According to the argument in the previous paragraph, this implies that it cannot be true that every $k$-fold covering of a finite set of points in ${\mathbb{R}}^2$ with straight lines splits into $\Omega(k)$ coverings. This consequence of Alon's theorem had been proved earlier, using the Hales-Jewett theorem  \cite{PaTT09}. Alon \cite{Al10} proved that the same example also disproves that all range spaces consisting of straight-line ranges in the plane admit $\eps$-nets of size $O(1/\eps)$.

\medskip
\noindent{\bf 4.} If in the proof of Theorems 1, 2, and 3, we replace Lemma 2.1 by the weaker inequality $|I|\le\frac{c^{d-1}}{\frac1{r-1}-\frac1{c-1}}$, established in \cite{PaT10} for every $c>r$, we obtain slightly weaker versions of Theorems~1, 2 and 3, with
$\Omega\left(\frac1\eps\log\frac1\eps/\log\log\frac1\eps\right)$ lower bounds on the sizes of the corresponding  $\eps$-nets. In a similar manner, if we replace Lemma~3.1 by a slightly weaker statement (Theorem~9) in \cite{ChPS09}, we obtain a weaker version of Theorem~4, with an
$\Omega\left(\frac1\eps\log\log\frac1\eps/\log\log\log\frac1\eps\right)$ bound on the size of the smallest $\eps$-net.

\bigskip
\noindent{\bf Acknowledgement.} We are very grateful to Boris Aronov and Micha Sharir for the many interesting discussions during the Special Semester on Discrete and Computational Geometry at EPFL in the Fall of 2010. Without their questions and remarks, this paper would have never been written.

\end{document}